\documentclass[preprint,12pt]{elsarticle}

\usepackage{amssymb}

\usepackage{amsmath}

\journal{Wave Motion}

\begin{document}

\begin{frontmatter}


 \cortext[cor1]{Corresponding author. \newline E-mail address: magdalini.koukouraki@espci.fr}

\title{Floquet scattering of shallow water waves by a vertically oscillating plate}

\affiliation[1]{organization={Laboratoire de Physique et Mécanique des Milieux Hétérogènes, ESPCI Paris,
PSL Research University, Sorbonne Université, Université Paris Cité},
             addressline={CNRS UMR 7636},
             city={Paris},
             postcode={75005},
             country={France}}
 \affiliation[2]{organization={Institut Langevin, ESPCI Paris, Université PSL},
             addressline={CNRS UMR 7587},
             city={Paris},
             postcode={75005},
             country={France}}
\affiliation[3]{organization={Laboratoire d’Acoustique de l’Université du Mans, Le Mans Université},
             addressline={CNRS UMR 6613},
             city={Le Mans},
             postcode={72085},
             country={France}}
\author{Magdalini Koukouraki $^{a,*}$, Philippe Petitjeans $^{a}$, Agnès Maurel $^{b}$ and Vincent Pagneux $^{c}$}

\begin{abstract}
We report on the scattering of a plane wave from a vertically oscillating plate in the low frequency approximation by means of Floquet theory. In the case of a static plate, the scattering coefficients are evaluated via mode matching method for the full two-dimensional linearised water wave problem and are compared with the coefficients obtained from a reduced one-dimensional model in the shallow water approximation. The main part of the analysis is the extension of this 1D shallow water approximation to the case of a vertically oscillating plate, where time modulation is only encapsulated in the blockage coefficient. We show that the incident wave is scattered into Floquet sidebands and extract the scattering coefficients for each harmonic using a Floquet scattering formalism. Finally, considering a slowly oscillating plate, we propose a quasistatic approximation which appears to be particularly accurate.
\end{abstract}

\begin{graphicalabstract}
\end{graphicalabstract}

\begin{highlights}
\item Floquet theory for the scattering of a plane wave by a vertically oscillating plate in shallow water.
\item The generation of harmonics and their characterisation when the plate is vertically oscillating.
\item A quasistatic approximation which is quite robust.
\end{highlights}


\end{frontmatter}

\section{Introduction}
Time-dependent systems involving wave-matter interactions have been extensively studied over the years, as they encompass intriguing wave phenomena which are universal from quantum mechanics to condensed matter and fluid mechanics \cite{tretyakov}, \cite{lurie_book}, \cite{monticone}, \cite{pena_opinion}. These effects include time reversal \cite{bacot}, frequency conversion \cite{apfell},  parametric amplification \cite{galiffi}, transient amplification \cite{kiorpe}, temporal waveguiding \cite{temporal_aiming} among many others. An increasing activity regarding time-varying media concerns electromagnetic platforms; it dates back to the pioneering works of Morgenthaler in the 1950s \cite{morgenthaler} on waves passing through media where the phase velocity is rapidly modified. Thenceforth, the research on time-varying metamaterials as a means to control and harness waves has significantly grown \cite{engheta_science}. Understanding the interaction of waves with scatterers exhibiting time-variation can be challenging and often requires the development of new theoretical and numerical tools. Floquet theory has been long applied for scattering problems in periodically driven systems, as discussed in \cite{reichl_1} and \cite{reichl_2} for the transmission of electrons through harmonically modulated potentials and in \cite{stefanou} for the propagation of waves in layered optomagnonic structures. 

In the realm of water waves, the bathymetry can play a key role in the wave dynamics and thus has been a topic of interest for many years in terms of wave scattering. More specifically, the reflection and transmission of waves by different bottom profiles has been already examined through the conformal mapping technique, which was first proposed by Fitz-Gerald (1976) \cite{fitz_gerald} and Hamilton (1977) \cite{hamilton}, and afterwards implemented by Evans and Linton (1994) \cite{evans_linton}. This technique relies on using a conformal transformation which enables to map an initial fluid domain with an irregular bottom boundary into a constant strip of fluid with the only added implication being an extra coefficient at the surface condition. Porter (2005) \cite{porter_map} subsequently revisited the technique to account for steep components in the bottom profile, concentrating on Roseau, shoaling and ridge-type profiles.

For the particular bathymetry with vertical barriers, numerous theoretical works have been made on wave scattering in infinitely-deep water \cite{newman}, \cite{porter_evans_vertical_barriers} and on structured beds composed of periodically arranged vertical plates in shallow water \cite{marangos_structured_bathymetry},  \cite{maurel_pham}, \cite{anisotropy}. Even though the forementioned literature involves static plates, there have also been works on moving underwater barriers and time-varying topographies. Evans (1970) in \cite{evans_vertical_plate} reports on the forces and moments on a vertical plate performing rolling oscillations under the water surface. Also, Tuck (1977) in \cite{tuck} treats the general scattering problem of water waves from a space-time dependent bottom profile in the shallow water approximation where the wavelength is much larger than the water depth, by using matched asymptotic expansions.

In this study, we focus on the linear long-wavelength water wave theory and we wish to model the interaction of a monochromatic wave with a submerged infinitely thin vertical plate, whose height is a function of time. First, we formulate the problem where the plate is static and then we proceed to the case where the plate is vertically oscillating. For the static plate, we present the full linearised problem and extract the two-dimensional field and the scattering coefficients for all range of frequencies. Then, by moving from the mid to the low frequency range of shallow water approximation, the field can be described effectively by the one-dimensional wave equation where the effect of the plate is incorporated in jump conditions at the plate position. For the vertically oscillating plate and in this low-frequency limit, we propose a Floquet theory approach in order to retrieve the reflected and transmitted fields, as well as a quasistatic approximation for the case where the plate is oscillating sufficiently slow compared to the period of the incident wave. Finally, we discuss the two methods in terms of their agreement and limitations.
\section{Scattering by a submerged vertical plate}
\subsection{Governing equations}
\begin{figure}[h]
\centering
\includegraphics[width=0.95\columnwidth]{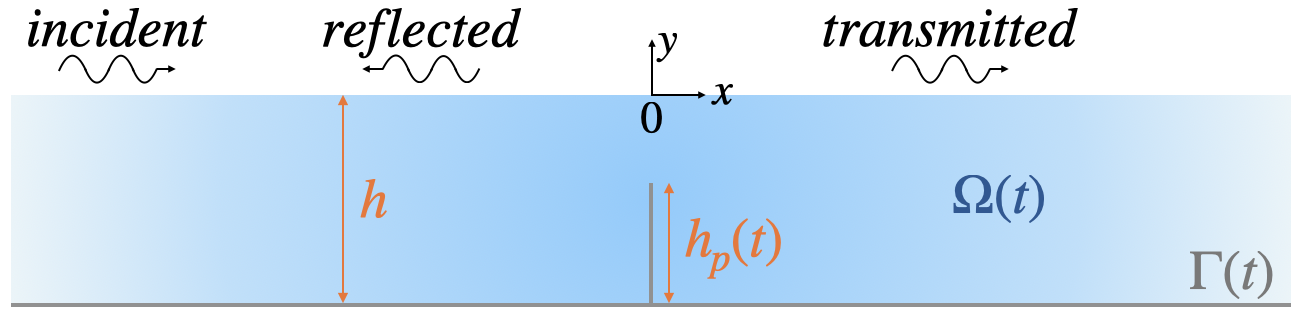}
\caption{Schematic representation of wave scattering by an infinitely thin plate of height $h_{p}(t)$ inside a channel of depth $h$.\label{fig1}}
\end{figure}

Let us consider an irrotational, incompressible and inviscid fluid of depth $h$, extending horizontally in an unbounded domain, and an infinitely thin plate of height $h_{p}$ sitting at the fluid bottom at position $x=0$, as depicted in Figure~\ref{fig1}.  We wish to characterize the scattering problem of a plane wave, incident on the plate from $x=-\infty$, when the plate height can also be allowed to vary with time. Following the classical linearised water-wave theory, which is adressed in one of the textbooks \cite{mei}, \cite{witham}, \cite{lamb}, \cite{stoker}, the problem translates as:
\begin{equation} \label{ww}
\left\{\begin{array}{l}
\Delta \Phi=0, \quad \text{in}  \quad \Omega(t) \\
\hat{\mathrm{n}} \cdot \nabla \Phi=0, \quad \text{on}  \quad \Gamma(t) \\
\frac{\partial \Phi}{\partial y}=-\frac{1}{g} \frac{\partial^{2}\Phi}{\partial t^{2}}, \quad y=0,
\end{array}\right.
\end{equation}
where $\Phi(x,y,t)$ denotes the velocity potential, $g$ the acceleration of gravity and $\hat{\mathrm{n}}$ the unit normal vector on $\Gamma(t)$. Note that by assuming that the plate has an infinitely small width, we guarantee that the plate will not act as a source when vertically oscillating (see Mei \cite{mei}). This is evident from the absence of a source term on the impermeable boundary condition on $\Gamma(t)$. 
\subsection{Shallow water model with jump conditions} \label{sw_section}
As demonstrated in \cite{tuck}, by starting from the system \eqref{ww} and implementing matched asymptotic expansions, one can eliminate the wave-field dependence on $y$ in the long-wavelength limit and obtain a reduced one-dimensional model with jump conditions at the position of the plate. This model reads as
\begin{subequations} \label{sw}
\begin{align}
\frac{\partial^2 \phi}{\partial x^2}-\frac{1}{c_{0}^{2}} \frac{\partial^2 \phi}{\partial t^2}=0, \label{eq:sw1}\\
{[\phi]_{0^{-}}^{0^{+}}=\left.2 B_\mu (t) h \partial_x \phi\right|_0, \quad\left[\partial_x \phi\right]_{0^{-}}^{0^{+}}=0}, \label{eq:sw2}
\end{align}
\end{subequations}
where $\phi(x,t)=\Phi(x,0,t)$ is now the $y$-independent velocity potential, $c_{0}=\sqrt{gh}$ is the velocity at which long waves propagate and $B_{\mu}$ is known as the blockage coefficient. $B_{\mu}$ is determined strictly from the geometrical profile of the fluid bottom and is a function of time when the topography is time-varying. For an infinitely thin plate, the blockage coefficient is given in the following explicit form:
\begin{equation} \label{blocage}
B_\mu(t)=-\frac{2}{\pi} \ln \left[\sin \left(\frac{\pi}{2}(1-\mu(t))\right)\right], \quad \mu=\frac{h_p(t)}{h},
\end{equation}
where $h_{p}(t)$ is the height of the oscillating vertical plate. Further information on the derivation of Equation \eqref{blocage} can be found in the paper \cite{porter_map}. \section{The static plate}
First, as a warm-up and in order to assess the validity range of the shallow water approximation, we are going to tackle the scattering problem for a static plate. 
\subsection{Mode matching method for any water depth regime}
In order to determine the reflection $R$ and transmission $T$ coefficients in the time-harmonic regime (convention $e^{-i\omega t}$) for each dimensionless frequency $\omega\sqrt{h/g}$ of the incident wave we follow the mode matching method. Although this method is already discussed in \cite{mei}, in this section we revisit its key points for our case, where the scatterer is an infinitely thin vertical plate. 

First, we rewrite the system \eqref{ww} in its time-independent form, by setting $\Phi=\Re\{\tilde{\Phi} e^{-i\omega t}\}$ and then dropping the tilde:
\begin{equation} \label{ww_harmonic}
\left\{\begin{array}{l}
\Delta \Phi=0, \quad \text{in}  \quad \Omega \\
\hat{\mathrm{n}} \cdot \nabla \Phi=0, \quad \text{on}  \quad \Gamma \\
\frac{\partial \Phi}{\partial y}=\frac{\omega^{2}}{g} \Phi, \quad y=0.
\end{array}\right.
\end{equation}
We start by expanding the solution (for $x \neq 0$) on the basis of orthonormal transverse eigenfunctions $g_{n}(y)$, which by solving the system \eqref{ww_harmonic} are found to be
\begin{equation}  \label{transverse_eig}
     g_{n}(y)=G_{n}\cosh[{k_{n}(y+h)}], \quad
     G_{n}=\sqrt{\frac{\sinh{(2k_{n}h})}{4k_{n}}+\frac{h}{2}},
\end{equation}
with $k_{n}$ the roots of the dispersion relation
\begin{equation}
     \omega^{2}=gk\tanh{kh}.
\end{equation}
The above functions satisfy both the Robin-type condition at $y=0$, and the Neumann boundary condition at $y=-h$. Therefore, the most general solution of the scattering problem reads as
\begin{equation} \label{sol_general_left}
    \Phi(x<0,y)=(e^{ik_{0}x}+Re^{-ik_{0}x})g_{0}(y)+\sum_{n=1}^{\infty}A_{n}e^{|k_{n}|x}g_{n}(y),
\end{equation}
\begin{equation} \label{sol_general_right}
    \Phi(x>0,y)=Te^{ik_{0}x}g_{0}(y)+\sum_{n=1}^{\infty}B_{n}e^{-|k_{n}|x}g_{n}(y),
\end{equation}
with $R$ and $T$ the reflection and transmission coefficients of the plane wave mode respectively and $A_{n}$ and $B_{n}$ are the coefficients of the evanescent modes which are excited near the plate. By looking at the symmetry of the problem, it is straightforward to split it into two sub-problems: a symmetric and an asymmetric part, 
\begin{equation}
     \Phi=\Phi_{s}+\Phi_{a}, \quad \text{with} \quad \Phi_{s}(x,y)=\Phi_{s}(-x,y), \quad \Phi_{a}(x,y)=-\Phi_{a}(-x,y),
\end{equation}
where we use the subscript "s" for symmetric and "a" for asymmetric. We then need to solve the problem just at the region $x<0$ and extend the solution for $x>0$. 

Due to symmetries, on the one hand, the boundary conditions which should be satisfied at $x=0$ are respectively 
\begin{equation} \label{bcasym}
    \frac{\partial \Phi_{s}}{\partial x}(x=0,S^{-})=\Phi_{a}(x=0,S^{-})=0,
\end{equation}
for the surface $S^{-}=\{x=0,y\in[-(h-h_p),0]\}$ above the plate.

On the other hand, the impermeability condition along the rigid surface of the plate $S_{p}=\{x=0,y\in[-h,-(h-h_{p})]\}$ yields
\begin{equation} \label{bcasym1}
    \frac{\partial \Phi_{s}}{\partial x}(x=0,S_{p})=\frac{\partial\Phi_{a}}{\partial x}(x=0,S_{p})=0.
\end{equation} 
Therefore, we use the following expansion for the left region with respect to the plate:
\begin{equation} \label{sol_sa_left}
    \Phi_{s,a}(x<0,y)=(e^{ik_{0}x}+R_{s,a}e^{-ik_{0}x})g_{0}(y)+\sum_{n=1}^{\infty}A_{n}e^{|k_{n}|x}g_{n}(y),
\end{equation}
where $R_{s,a}$ denote the reflection coefficients of each subproblem and it can be shown that
\begin{equation}
R=\frac{1}{2}\left(R_{s}+R_{a}\right), \quad T=\frac{1}{2}\left(R_{s}-R_{a}\right),
\end{equation}
with $|R_{s,a}|=1$ because of energy conservation. The symmetric part has a trivial solution, since the Neumann boundary condition at $S=S_{p}\bigcup S^{-}$ yield $R_{s}=1$. Hence, the only part which contributes to the variation of the scattering coefficients with frequency is $R_{a}$, the asymmetric one. 

\begin{figure}[h]
\centering
\includegraphics[width=1\columnwidth]{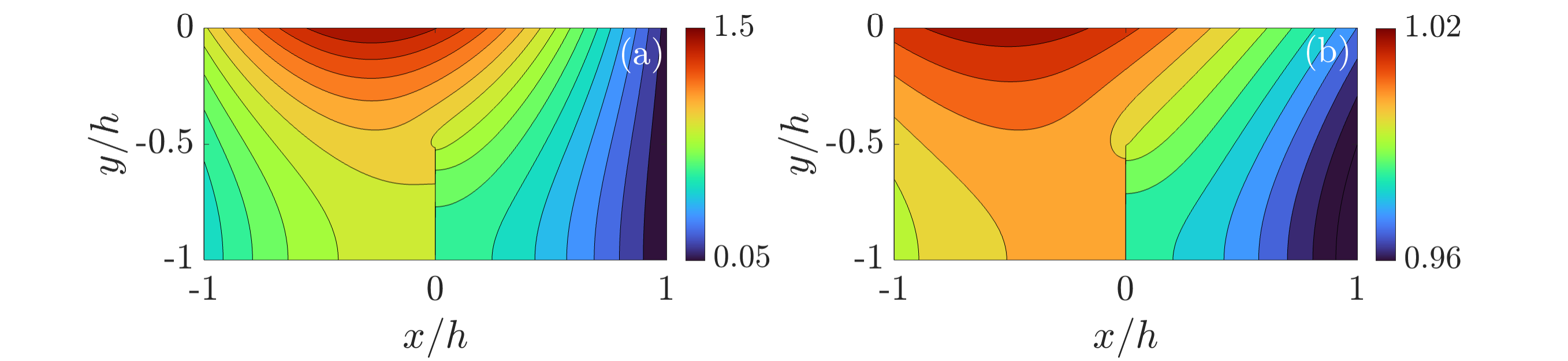}
\caption{Two-dimensional profile of the wave field recovered via mode-matching method for (\textbf{a}) $\mu=0.5$, $\omega\sqrt{h/g}=1$ and (\textbf{b}) $\mu=0.5$, $\omega\sqrt{h/g}=0.2$.\label{fig2}}
\end{figure}

It is convenient to write the asymmetric part of solution \eqref{sol_sa_left} in a more general form, such as 
\begin{equation} \label{series_a_x_0}
    \Phi_{a}(x\leq0,y)=\sum_{n=0}^{\infty}a_{n}(x)g_{n}(y),
\end{equation}
where we decompose the coefficients $a_{n}$ into the ones of the incoming wave and of the reflected wave,
\begin{equation} \label{a_n_whole}
a_{n}(x)=a_{n,0}(x)+a_{n,r}(x),
\end{equation}
with $a_{n,0}(x)=\delta_{0n}e^{ik_{0}x}$ and $a_{n,r}(x)=C_{n}e^{-ik_{n}x}$. Next, we apply condition \eqref{bcasym} for the component $\Phi_{a}(x=0,S^{-})$, and project over the transverse functions
\begin{equation}  \label{transverse_eig_hat}
     \hat{g}_{p}(y)=\hat{G}_p\cosh{[\hat{k}_{p}(y+(h-h_p))]}, \quad
     \hat{G}_{p}=\sqrt{\frac{\sinh{(2\hat{k}_{p}(h-h_p)})}{4\hat{k}_{p}}+\frac{(h-h_p)}{2}},
\end{equation}
with $\hat{k}_{p}$ satisfying $\omega^{2}=g\hat{k}\tanh{\hat{k}(h-h_{p})}.$ Doing so we obtain
\begin{equation} \label{condition1}
 \mathrm{F^{t}}\vec{a}(0)=0,
\end{equation}
with the vector $\vec{a}(0)$ containing the components $a_{n}(x=0)$ and the matrix elements of $\mathrm{F}$ are given as
 \begin{equation}
 (\mathrm{F})_{m p}=\int_{-(h-h_{p})}^{0}g_{m}(y)\hat{g}_{p}(y)dy.
\end{equation}
Then, we can rewrite the continuity condition of $\partial_{x}\Phi_{a}$ at $x=0$ as
\begin{equation} \label{der_sol_x_0}
\partial_{x}\Phi_{a}|_{x=0}=
\left\{ \begin{array}{l}
0, \quad \text{at} \quad S_{p}\\
\displaystyle \sum_{p=0}^{\infty}c_{p}\hat{g}_{p}(y), \quad \text{at} \quad S^{-}.
\end{array}\right.
\end{equation}
Deriving Equation \eqref{series_a_x_0} with respect to $x$, using Equation \eqref{der_sol_x_0}, and projecting on the functions $g_{m}$ we have that 
 \begin{equation} \label{condition2}
\vec{a}^{'}(0)=\mathrm{F} \vec{c}.
\end{equation}
Taking the derivative of $a_{n}$ in Equation \eqref{a_n_whole}, we have
\begin{equation} \label{a_dot}
\vec{a}^{'}(0)=\mathrm{Y}(\vec{a_{0}}(0)-\vec{a_{r}}(0)),
\end{equation}
where $(a_{0})_{m}=\delta_{m0}$ and $\mathrm{Y}$ a diagonal matrix with elements $(\mathrm{Y})_{m,m^{'}}=ik_{m}\delta_{m,m^{'}}$.
Next, a simple manipulation of Equations \eqref{condition1} and \eqref{a_n_whole} yields the relation
\begin{equation}\label{unknown}
    \mathrm{F^{t}}\vec{a_{r}}(0)=-\mathrm{F^{t}}\vec{a_{0}}(0).
\end{equation}
Finally, combining Equations \eqref{unknown}, \eqref{a_dot} and \eqref{condition2} one can extract that
\begin{equation}\label{4}
\vec{a_{r}}(0)=\vec{a_{0}}(0)-\mathrm{Y^{-1} F}\vec{c},
\end{equation}
where
\begin{equation}\label{3}
\vec{c}=2\mathrm{(F^{t}YF)^{-1}F^{t}}\vec{a_{0}}(0).
\end{equation}
Computing Equations \eqref{3} and \eqref{4} numerically and using a number of modes around $N=50$ for the expansion on $S (N,g_{n})$ and a number $P\sim (S^{-}/S)N$ for the expansion on $S^{-} (P,\hat{g}_{p})$, we retrieve $R$ and $T$ as well as the full wave field for each frequency. In Figure~\ref{fig2} we show the two-dimensional field recovered for ratio $\mu=0.5$ and two water depth regimes: the finite-water depth and the shallow water limit. As illustrated, the wave field becomes almost homogeneous with $y$ when $\omega \sqrt{h/g}\ll 1$ (notice the variation of the colorbar on panel (b)), which justifies simplifying the set of equations \eqref{ww_harmonic} to just one partial differential equation at the surface supplemented with the effective boundary conditions at $x=0$. 
\subsection{Low frequency approximation: the shallow water regime}\label{low_freq_approx_static}
As was previously discussed, for waves of sufficiently small amplitude (linear regime) and of very long wavelength compared to the water depth (shallow water regime), the wave dynamics can be well approximated by the wave equation \eqref{eq:sw1}. In this shallow water approximation (SWA), waves are nondispersive and satisfy the dispersion relation $\omega=c_{0}k$, with $k$ the wave number. 

Considering once again the harmonic regime, so that $\phi=\Re\{f(x) e^{-i\omega t}\}$, and a non-moving plate, such that $B_{\mu}(t)=\mathrm{const}$ in Equation \eqref{eq:sw2}, we construct the solution of Equation \eqref{eq:sw1} as follows: For $x<0$, the wave is composed of an incident, right-propagating, wave and a reflected wave, $f^{-}=e^{ikx}+R_{sw}e^{-ikx}$, while for $x>0$ the solution takes the form of a transmitted wave, $f^{+}=T_{sw}e^{ikx}$. By using these expressions for $f^{+}$ and $f^{-}$ when applying the continuity of $\partial_{x}\phi$ and the jump condition of $\phi$ at $x=0$ (Equation \eqref{eq:sw2}), we obtain the reflection and transmission coefficient in the shallow water approximation as
\begin{equation} \label{RT_sw}
R_{sw}  =-\frac{i k h B_\mu}{1-i k h B_\mu}, \quad T_{sw} =\frac{1}{1-i k h B_\mu}.
\end{equation}
Notice that these scattering coefficients satisfy the energy conservation $|R_{sw}|^{2}+|T_{sw}|^{2}=1$. 

\begin{figure}[h]
\centering
\includegraphics[width=0.95\columnwidth]{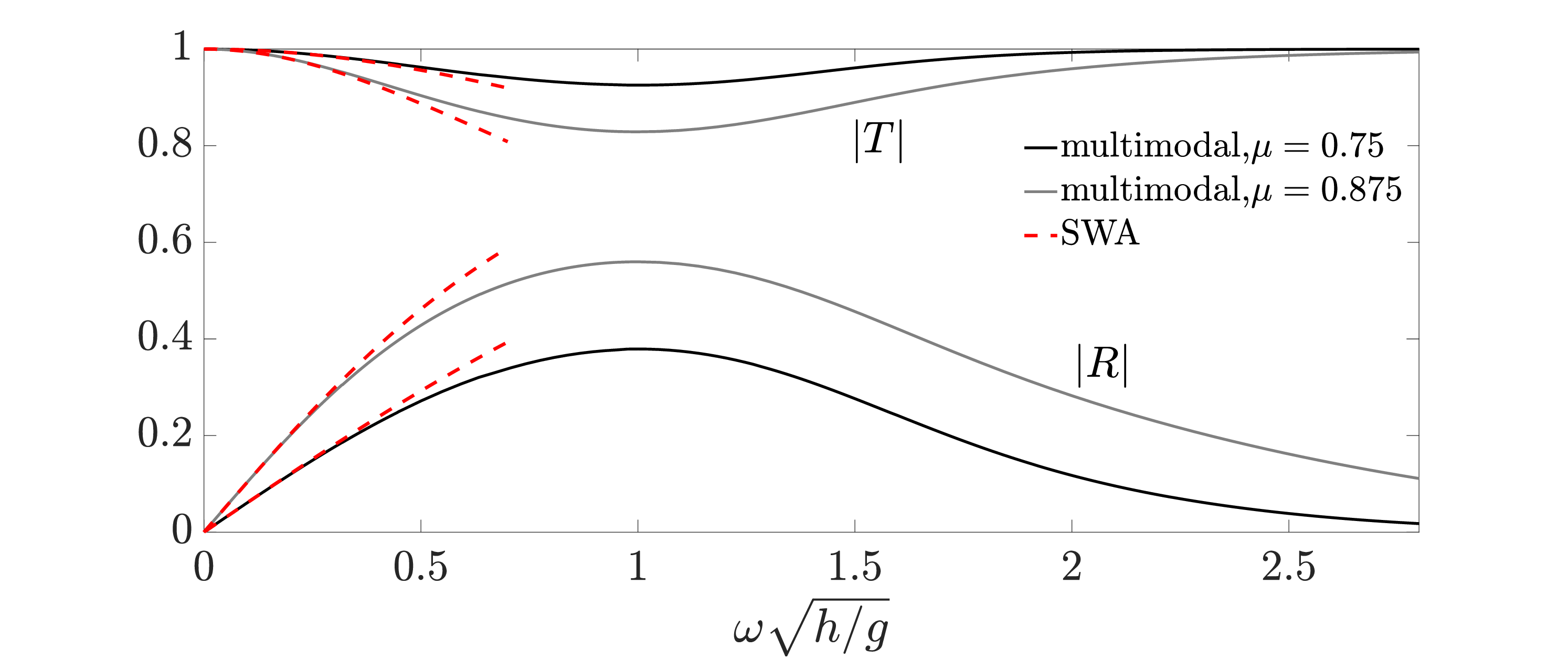}
\caption{Variation of $|R|$ and $|T|$ for $\mu=0.75$ and $\mu=0.875$. The shallow water approximation (SWA) is given by Equation \eqref{RT_sw}.\label{fig3}}
\end{figure}

In Figure~\ref{fig3}, we portray $R$ and $T$ obtained from the mode matching method along with the relations \eqref{RT_sw} with respect to the frequency for two values of $\mu$. Notice that the reflection increases as the plate height approaches the surface (increasing $\mu$). Furthermore, the SWA holds sufficiently up to $\omega \sqrt{h/g}=0.2$, which leads us to establish a frequency interval of $\omega \sqrt{h/g} \in [0, 0.5]$ for our study in the shallow water limit.
\section{The vertically oscillating plate: shallow water regime}
Now, we are  going to the main part of this work where we evaluate the scattering by
a vertically oscillating plate in the shallow water approximation.
\subsection{Floquet theory} \label{Floquet theory}
In this paper we are interested in the simplest case where the time variation is introduced in the blockage coefficient. We hereafter consider that 
\begin{equation} \label{blocage_TV}
B_{\mu}(t)=B_{\mu,0}+B_{\mu,1}\cos(\omega_{p}t),
\end{equation}
 with $\omega_{p}$ the characteristic parameter of oscillation, which consequently leads to a  more complex plate motion, defined as
\begin{equation}
\mu(t)=1-\frac{2}{\pi} \operatorname{asin}\left[\exp \left(-\frac{\pi}{2} B_\mu(t)\right)\right].
\end{equation}
Our starting point is the problem already introduced in Section~\ref{sw_section} with a right-propagating monochromatic wave of frequency $\omega$ impinging on the periodically driven scatterer. First, Floquet theorem allows us to write the solution in the form
\begin{equation}\label{floquet}
\phi=\Re\{e^{-i \omega t} \psi\}, \quad \psi(x,t)=\psi(x,t+T_{p}),
\end{equation}
with $T_{p}=2\pi/\omega_{p}$. Since $\psi$ is periodic, it can be expanded in the Fourier series
\begin{equation}\label{fourier}
\psi(x,t)=\sum_{n}\psi_{n}(x)e^{-in\omega_{p}t},
\end{equation}
with  $n\in(- \infty, + \infty)$ and the Fourier modes $e^{-in\omega_{p}t}$ satisfying the orthogonality relation:
\begin{equation} \label{orthogonality}
\frac{1}{T_{p}}\int_{0}^{T_{p}}e^{i(m-n)\omega_{p}t}dt=\delta_{mn}.
\end{equation}
Plugging Equations \eqref{floquet} and \eqref{fourier} into Equation \eqref{eq:sw1}, projecting on $e^{-in\omega_{p}t}$ and using Equation \eqref{orthogonality}, we find for $x\neq0$ that 
\begin{equation}\label{Helmholtz_psin}
\frac{d^2 \psi_n}{d x^2}+k_n^2 \psi_n=0, \quad k_n=\frac{\omega_{n}}{c_0},
\end{equation}
where $\omega_{n}=\omega+n \omega_p$. It follows from Eq. \ref{eq:sw2} that the boundary condition for $\psi_n$ reads as:
\begin{equation}\label{psi_n_jump}
[\psi_{n}^{\prime}]_{0^{-}}^{0^{+}}=0, \quad [\psi_{n}]_{0^{-}}^{0^{+}}=2B_{\mu,0}h\psi_{n}^{\prime}+B_{\mu,1}h(\psi_{n+1}^{\prime}+\psi_{n-1}^{\prime}).
\end{equation}
Hence, one can write the solution for $\psi_{n}$ in the regions $x<0$ and $x>0$ as follows:
\begin{equation}\label{psi_n_minus}
\psi_{n}^{-}(x<0)=\delta_{0n}e^{ik_{n}x}+r_{n}e^{-ik_{n}x}, 
\end{equation}
\begin{equation}\label{psi_n_plus}
\psi_{n}^{+}(x>0)=t_{n}e^{ik_{n}x},
\end{equation}
with $r_{n}$ and $t_{n}$ the scattering coefficients of each harmonic. The next step is to evaluate the jump conditions by substituting Equations \eqref{psi_n_minus} and \eqref{psi_n_plus} in Equation \eqref{eq:sw2}, in order to derive the expressions for $r_{n}$ and $t_{n}$. First, we proceed with the continuity of $\partial_{x}\phi$ at $x=0$, which yields
\begin{equation}
\sum_n (\delta_{0n}-r_n) k_n e^{-i n \omega_p t}=\sum_n t_n k_n e^{-i n \omega_p t}.
\end{equation}
By projecting on the Fourier modes, using the orthogonality relation \eqref{orthogonality}, and representing the scattering coefficients in vector forms, with vector components $(t)_{m}=t_{m}$, $(r)_{m}=r_{m}$, we find that
\begin{equation} \label{1}
\vec{t}=\vec{b}-\vec{r},
\end{equation}
where $(b)_{m}=\delta_{0m}$.  Notice that $r_{m}=-t_{m}$, for $m \neq 0$, which is consistent with the symmetry of the problem.

Then, the discontinuity of $\phi$ at $x=0$ translates into
\begin{equation}
\begin{gathered}
\sum_n\left(t_n-\delta_{0n}-r_n\right) e^{-i n \omega_p t}= \\
\sum_{n}ik_n \left( \delta_{0n}-r_{n}\right) \left[ 2  B_{\mu, 0} h e^{-i n \omega_p t}+ B_{\mu, 1} h \left(e^{-i(n-1) \omega_p t}+e^{-i(n+1) \omega_p t}\right)\right],
\end{gathered}
\end{equation}
which after following the same procedure as before can be adapted into a form of a linear system:
\begin{equation}\label{eq36}
    \vec{t}=(\mathrm{I}+\mathrm{V})\vec{b}+(\mathrm{I}-\mathrm{V})\vec{r},
\end{equation}
with
\begin{equation}
(\mathrm{V})_{m, m^{\prime}}=i B_{\mu, 1} h \left(k_{m^{\prime}+1}\delta_{m, m^{\prime}+1}+k_{m^{\prime}-1}\delta_{m, m^{\prime}-1}\right)+2i B_{\mu, 0} h k_{m^{\prime}}\delta_{m, m^{\prime}},
\end{equation}
and $\mathrm{I}$ denoting the identity matrix. Since $\mathrm{V}$ induces coupling between the harmonics, Equation \eqref{eq36} demonstrates the fact that the incident wave is scattered into Floquet sidebands with frequencies $\omega_{n}$. Plugging Equation \eqref{1} into Equation \eqref{eq36} we obtain the relation 
\begin{equation} \label{2}
\vec{r}=-(2\mathrm{I}-\mathrm{V})^{-1} \mathrm{V} \vec{b}.
\end{equation}
Finally, by performing some algebraic manipulations combining the boundary conditions \eqref{psi_n_jump} for $\psi_{n}$ and the form of solutions for $\psi_n$ given in Equations \eqref{psi_n_minus} and \eqref{psi_n_plus}, one can derive the conservation of the quantity
\begin{equation} \label{conservation}
\mathcal{T}+\mathcal{R}=1,
\end{equation}
with $\mathcal{T}=\sum_{n}|t_{n}|^{2}k_{n}/k_{0}$, and  $\mathcal{R}=\sum_{n}|r_{n}|^{2}k_{n}/k_{0}$. More precisely, for this derivation one needs to proceed as follows. First, one evaluates the discontinuity of the product $\bar{\psi}_{n} \psi_{n}^{\prime}$ at $x=0$ using Equation \eqref{psi_n_jump}:
\begin{equation} 
    [\bar{\psi}_{n} \psi_{n}^{\prime}]_{0^{-}}^{0+}=\left[2B_{\mu,0}h \bar{\psi}_{n}^{\prime}(0)+B_{\mu,1}h\left(\bar{\psi}_{n+1}^{\prime}(0)+\bar{\psi}_{n-1}^{\prime}(0)\right)\right]\psi_{n}^{\prime}(0).
\end{equation}
Then, one takes the sum of this product over all $n$-harmonics which gives
\begin{equation} \label{sum_psi_n_psi_n_dot}
    [\sum_{n}\bar{\psi}_{n} \psi_{n}^{\prime}]_{0^{-}}^{0+}=2B_{\mu,0}h \sum_{n}|\psi_{n}^{\prime}(0)|^{2}+B_{\mu,1}h\sum_{n} \left(\psi_{n}^{\prime}(0)\bar{\psi}_{n+1}^{\prime}(0)+\bar{\psi}_{n-1}^{\prime}(0)\psi_{n}^{\prime}(0)\right).
\end{equation}
Rewriting that $\sum_{n}\bar{\psi}_{n-1}^{\prime}\psi_{n}^{\prime}=\sum_{n}\bar{\psi}_{n}^{\prime}\psi_{n+1}^{\prime}+\bar{\psi}_{-N-1}^{\prime}\psi_{-N}^{\prime}-\bar{\psi}_{N}^{\prime}\psi_{N+1}^{\prime}$ and given than $\psi_{-N-1}=\psi_{N+1}=0$, since $n \in [-N,N]$, Equation \eqref{sum_psi_n_psi_n_dot} takes the form
\begin{equation}
    [\sum_{n}\bar{\psi}_{n} \psi_{n}^{\prime}]_{0^{-}}^{0+}=2B_{\mu,0}h \sum_{n}|\psi_{n}^{\prime}(0)|^{2}+2B_{\mu,1}h\sum_{n}\Re\{\psi_{n}^{\prime}(0)\bar{\psi}_{n+1}^{\prime}(0)\}.
\end{equation}
From there one finds that
\begin{equation}
    [\Im\{\sum_{n}\bar{\psi}_{n} \psi_{n}^{\prime}\}]_{0^{-}}^{0+}=0,
\end{equation}
which indicates the conservation of the total flux at $x=0$. Furthermore, since the total flux is also conserved for $x<0$ and $x>0$ separately, a property stemming from Equation \eqref{Helmholtz_psin}, it is conserved everywhere in space: $[\Im\{\sum_{n}\bar{\psi}_{n} \psi_{n}^{\prime}\}]_{-\infty}^{+\infty}=0$. In the end one obtains Equation \eqref{conservation}. Incidentally, this conservation law is also found in the case of the Schrödinger equation (see \cite{stefanou}), referring to the probability current conservation, with $\mathcal{T}$ the transmittance and $\mathcal{R}$ the reflectance.  
\subsubsection{Generation of harmonics} \label{section_harmonic_generation}
\begin{figure}[h] 
\centering
\includegraphics[width=6.75cm]{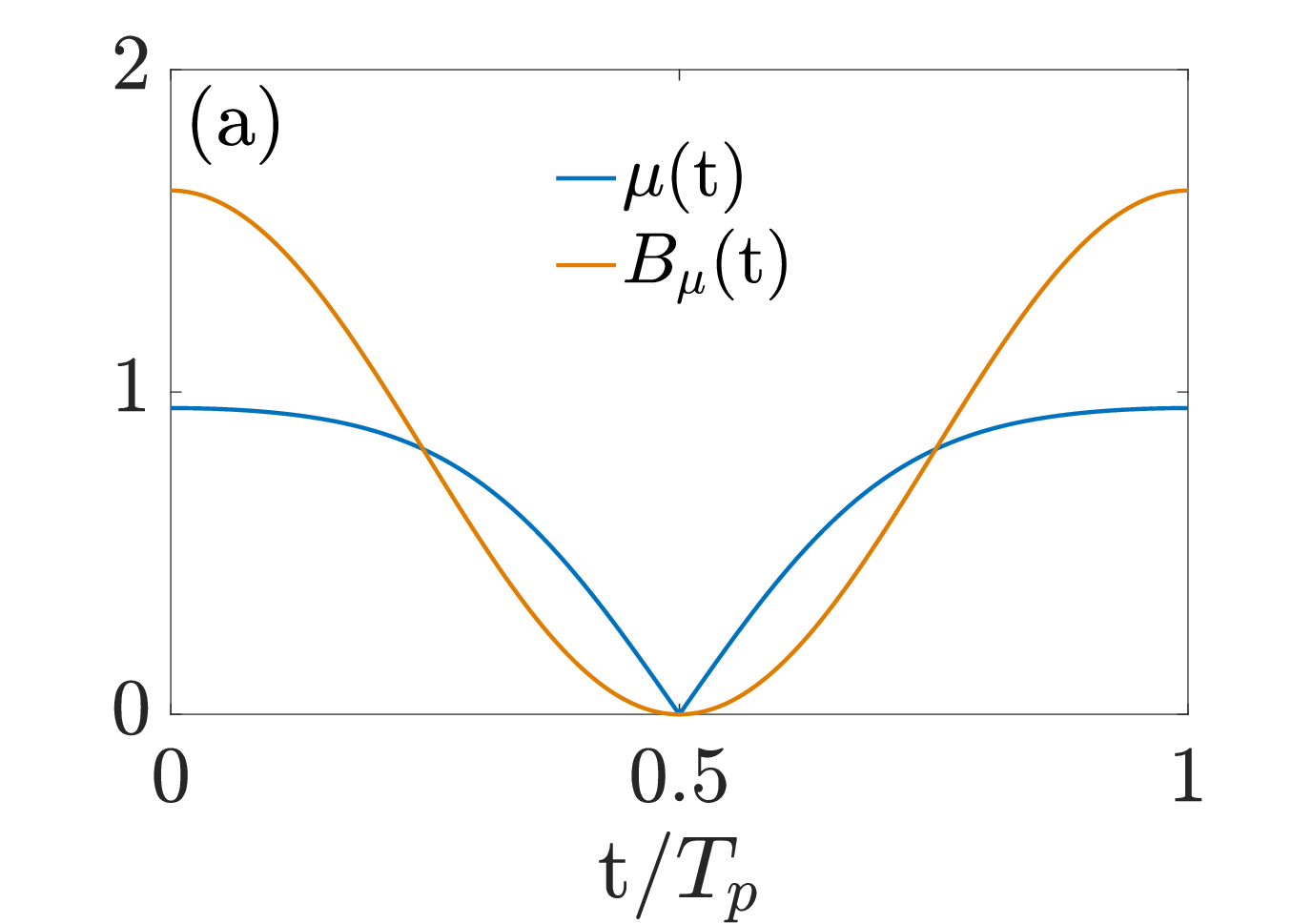}
\includegraphics[width=6.75cm]{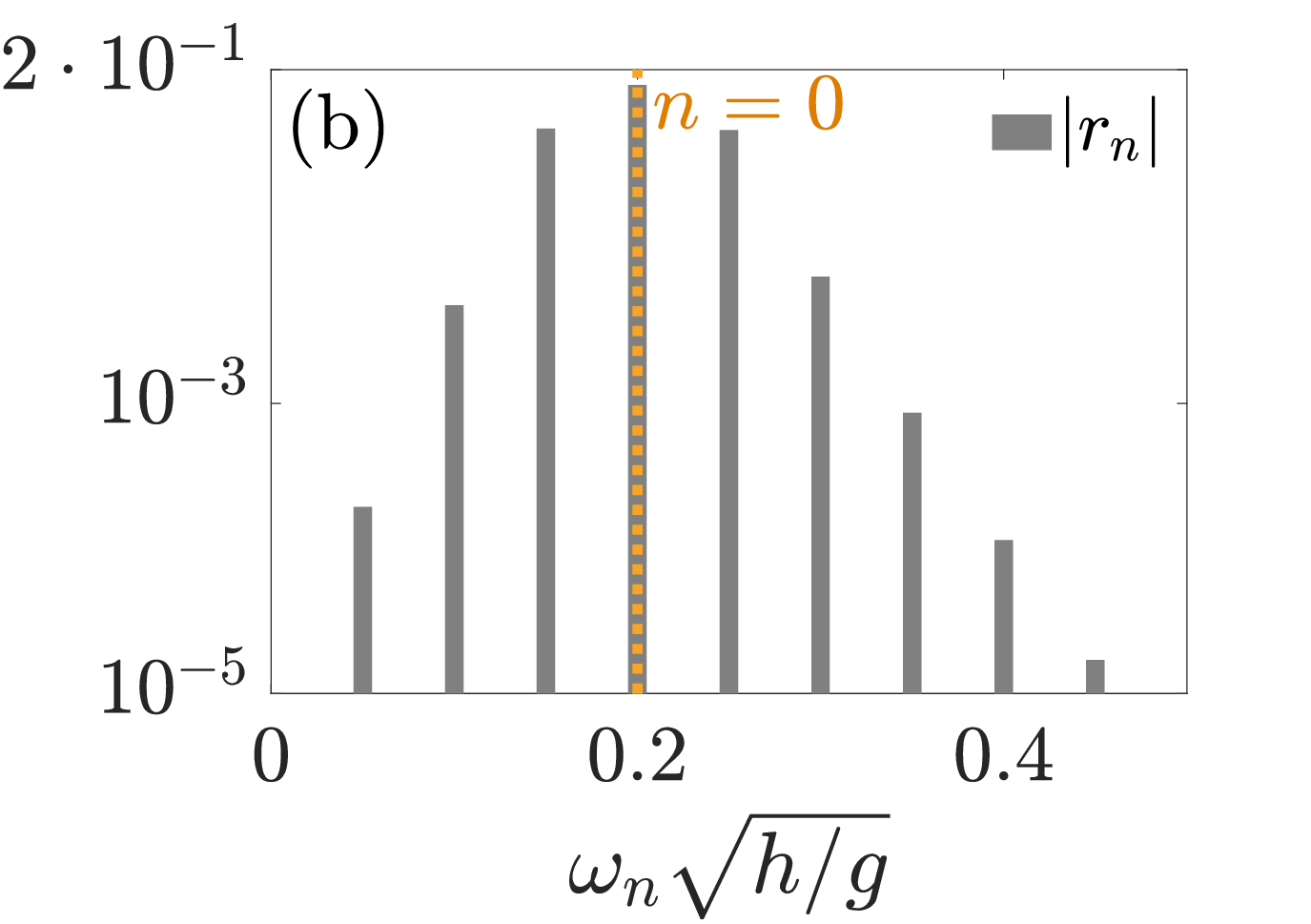}
\caption{(\textbf{a}) Temporal variation of the blockage coefficient in one period $T_{p}$, and the corresponding plate oscillation. (\textbf{b}) Reflection coefficients of the generated harmonics for the plate oscillation of panel (a), with $\omega\sqrt{h/g}=0.2$ and $\omega_{p}=\omega/4$.\label{fig4}}
\end{figure}

Equipped with Equations \eqref{1} and \eqref{2}, we can now investigate and quantify the harmonic generation. In order for the plate to have a visible impact on the wave at low frequencies, we choose to impose a strong vertical plate movement as shown in Figure~\ref{fig4}a, where $\mu(t) \in [0,0.95]$. For this configuration, and setting $\omega\sqrt{h/g}=0.2$ and $\omega_{p}=\omega/4$, the harmonics which are produced at $n=\pm 1$ represent roughly $49\%$ of the fundamental one in terms of reflection, while at each higher order of $n$ the coefficient $|r_{n}|$ drops an order of magnitude (see Figure~\ref{fig4}b). While in this example we witness an important contribution of the first sidebands ($n=\pm1$) in the total reflected field, we wish to see if this is still true for different frequencies of oscillation. For this purpose we fix the frequency of the incident wave and vary $\omega_{p}$, so as to uncover how each harmonic is affected by this variation.

\begin{figure}[h]
\centering
\includegraphics[width=6.75cm]{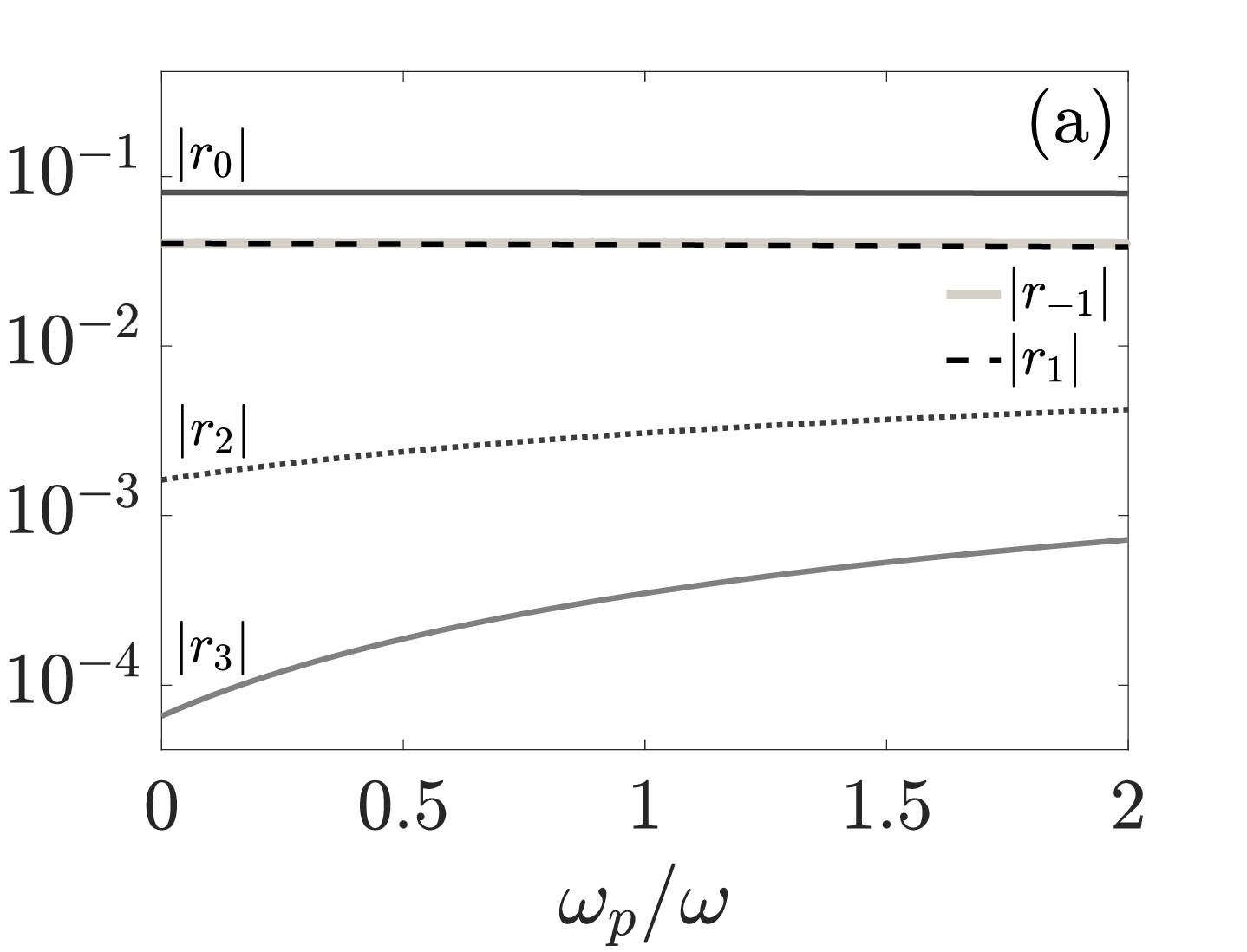}
\includegraphics[width=6.75cm]{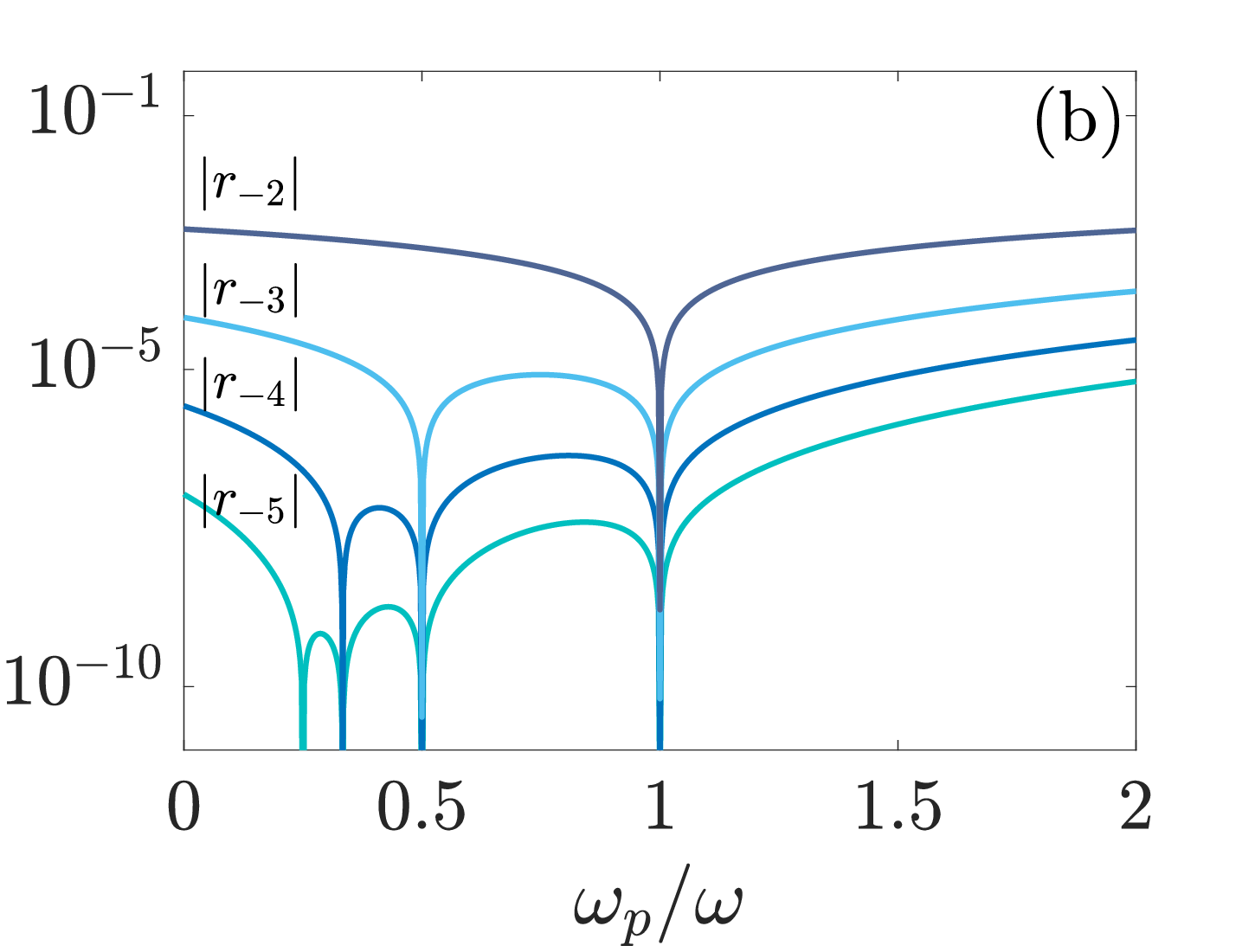}
\caption{Variation of the reflection coefficient of each harmonic in terms of $\omega_{p}$ normalized by the fixed incident frequency $\omega \sqrt{h/g}=0.1$, for $-1\leq n \leq 3$ in (\textbf{a}), and for $-5 \leq n \leq -2$ in (\textbf{b}).\label{fig5}}
\end{figure}

In Figure~\ref{fig5} we depict a numerical application, where $\omega\sqrt{h/g}=0.1$ and $\omega_{p} \sqrt{h/g} \in [0,0.5]$, plotting the harmonics corresponding to $-1 \leq n \leq 3$ on panel (a) and $-5 \leq n \leq -2$ on panel (b). It is clear that the harmonics with $n\geq -1$ are monotonous with $\omega_{p}/\omega$, and that the dependence on the plate oscillation decreases as we move from $n=3$ towards the fundamental harmonic, which appears constant compared to its counterparts. The first sidebands at $n=\pm1$ are almost identical, and we find that $|r_{-1}|\simeq |r_{1}|$ as $\omega_{p}/\omega\rightarrow 0$, which will be explained in the following section concentrating on the quasistatic adiabatic limit. Interestingly, on panel (b) we observe the elimination of all $r_{n}$ with $n<-1$ at $\omega=\omega_{p}$ and the elimination of all $r_{n}$ with $n<-2$ at $\omega=2\omega_{p}$. 
In fact, it appears that 
\begin{equation}
    |r_{n}|=0, \quad \text{for} \quad n<-m, \quad \text{when} \quad \omega=m\omega_{p},
\end{equation}
with $m\in \mathrm{N^{*}}$, $n\in \mathrm{Z}$. Indeed, at these frequencies the system \eqref{2} decouples and explicit relations can be found for $n\geq-m$. For instance, for $m=1$ one can extract that
\begin{equation}
    r_{-1}=\frac{iB_{\mu,1}hk_{0}}{2}(r_{0}-1),
\end{equation}
\begin{equation}
    r_{1}=\frac{iB_{\mu,0}hk_{0}+r_{0}\left(1-iB_{\mu,0}hk_{0} \right)}{iB_{\mu,1}hk_{0}},
\end{equation}
\small
\begin{equation}
    r_{2}=-\frac{\left\{2iB_{\mu,0}hk_{0}+(4B_{\mu,0}^{2}-B_{\mu,1}^{2})h^{2}k_{0}^{2}+[2-6iB_{\mu,0}hk_{0}-(4B_{\mu,0}^{2}-B_{\mu,1}^{2})h^{2}k_{0}^{2}]r_{0}\right\}}{3B_{\mu,1}^{2}h^{2}k_{0}^{2}},
\end{equation}
\normalsize
and, as an extension, one can deduce that $r_{n}=s_{n}(r_{0})$, with $n\geq -1$ and $s_{n}$ the allocated function. This property of the system means that there are no reflected waves of the form $e^{i|k_{n}|(x+c_{0}t)}$. However, for a non integer ratio of $\omega_{p}/\omega$ this is no longer the case and it leads to harmonics with a reversed wave phase; it is visible from the rapid increase of the harmonic $|r_{-2}|$ from a practically zero value at $\omega_{p}/\omega=1$ to the order of $10^{-4}$ when $\omega_{p}/\omega=1.1$. It is also worth commenting on the fact that while in the Schrödinger equation the harmonics with $\omega_{n}<0$ result in evanescent modes (see \cite{reichl_1}, \cite{stefanou}), for the wave equation there is no restriction of that matter.   
\subsection{Quasistatic approximation}
While Floquet theory gives a direct insight into the reproduced harmonics and their dependence on the frequency of oscillation, we have seen that we have small changes in the $n$-harmonic amplitude with $\omega_{p}$ for $n=-1,0,1$, implying the existence of a quasistatic (QS) regime ($\omega_{p}\rightarrow 0$) that needs to be examined.

\begin{figure}[h]
\centering
\includegraphics[width=0.99\columnwidth]{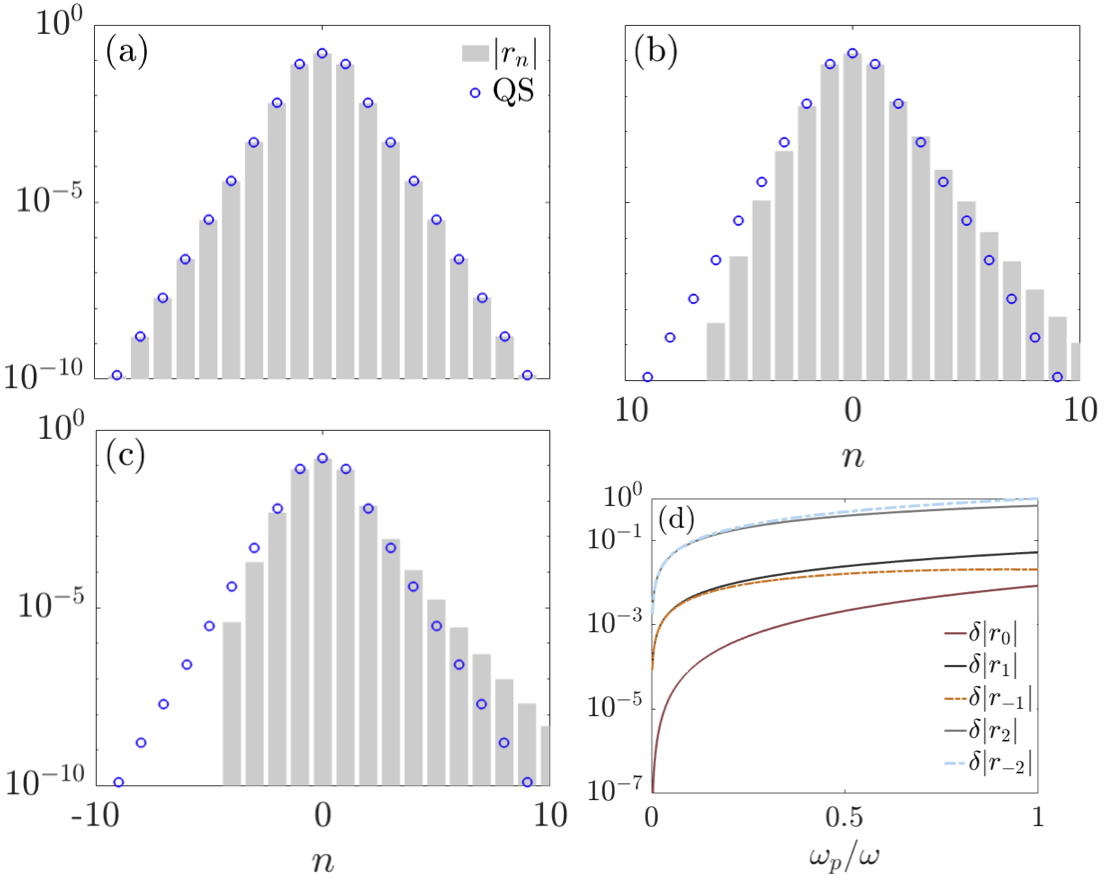}
\caption{Reflection amplitudes calculated by the Floquet theory (in gray bars) and by the QS approximation (in blue points) for $\omega \sqrt{h/g}=0.1$, with (\textbf{a}) $\omega_{p}\sqrt{h/g}=\omega/1000$, (\textbf{b}) $\omega_{p}\sqrt{h/g}=\omega/6$ and (\textbf{c}) $\omega_{p}\sqrt{h/g}=\omega/4$. (\textbf{d}) Relative difference of the two methods in terms of $\omega_{p}/\omega$ for the harmonics of indexes $-2\leq n\leq 2$.\label{fig6}}
\end{figure}

In order to understand this quasistatic adiabatic limit, we focus on a barrier moving much slower than the period of the incident wave, such that $\omega_{p}\ll \omega$. Then, the static solution for the reflected and the transmitted waves given in Section~\ref{low_freq_approx_static} is modified only by adjusting the time-dependent blockage coefficient (Equation \eqref{blocage_TV}) in the relations \eqref{RT_sw}. Hence, this yields
\begin{equation}\label{r_qs}
    \tilde{f}_{r}(x,t)=-\frac{ikh(B_{\mu,0}+B_{\mu,1}\cos(\omega_{p}t))}{1-ikh(B_{\mu,0}+B_{\mu,1}\cos(\omega_{p}t))}e^{-i(kx+\omega t)},
\end{equation}
\begin{equation}\label{t_qs}
    \tilde{f}_{t}(x,t)=\frac{1}{1-ikh(B_{\mu,0}+B_{\mu,1}\cos(\omega_{p}t))}e^{i(kx-\omega t)},
\end{equation}
with $\tilde{f}_{r}$ and $\tilde{f}_{t}$ denoting the reflected and transmitted wave in this adiabatic limit. Next, we Fourier expand the two components using the series
\begin{equation}\label{fourier_r_qs}
    \tilde{f}_{r}=\sum_{n}\tilde{r}_{n}e^{i(kx-\omega_{n}t)},
\end{equation}
\begin{equation}\label{fourier_t_qs}
    \tilde{f}_{t}=\sum_{n}\tilde{t}_{n}e^{i(kx-\omega_{n}t)},
\end{equation}
with $\omega_{n}=\omega+n\omega_{p}$. Combining Equations \eqref{r_qs}, \eqref{t_qs} with Equations \eqref{fourier_r_qs}, \eqref{fourier_t_qs}, then projecting on $e^{-in\omega_{p}t}$ and using the orthogonality of the modes, one can express the coefficients $\tilde{r}_{n}$ and $\tilde{t}_{n}$ in the quasistatic approximation as
\begin{equation} \label{int1}
    \tilde{r}_{n}=\frac{1}{T_{p}}\int^{T_{p}}_{0}\frac{ikh(B_{\mu,0}+B_{\mu,1}\cos(\omega_{p}t))}{ikh(B_{\mu,0}+B_{\mu,1}\cos(\omega_{p}t))-1}e^{-in\omega_{p}t}dt,
\end{equation}
\begin{equation} \label{int2}
    \tilde{t}_{n}=\frac{1}{T_{p}}\int^{T_{p}}_{0}\frac{1}{1-ikh(B_{\mu,0}+B_{\mu,1}\cos(\omega_{p}t))}e^{in\omega_{p}t}dt.
\end{equation}
By applying contour integration we obtain the expressions
\begin{equation}\label{t_QS_explicit}
    \tilde{t}_{n}=\frac{\left(\beta/\gamma -\sqrt{(\beta/\gamma)^{2}-1} \right)^{-|n|}}{\sqrt{\beta^{2}-\gamma^{2}}},
\end{equation}
\begin{equation}\label{r_QS_explicit}
    \tilde{r}_{n}=\delta_{0n}-\frac{\left(\beta/\gamma -\sqrt{(\beta/\gamma)^{2}-1} \right)^{-|n|}}{\sqrt{\beta^{2}-\gamma^{2}}},
\end{equation}
with $\beta=1-ikhB_{\mu,0}$ and $\gamma=ikhB_{\mu,1}$. Notice that the coefficients are symmetric around the fundamental, i.e. $\tilde{r}_{n}=\tilde{r}_{-n}$ for $n\neq 0$, which is not the case in the Floquet theory (see Figure~\ref{fig6}b,c). They are also independent from $\omega_{p}$, as expected.

In order to test the robustness and the limitations of the QS approximation, as opposed to the Floquet theory, we once again fix the incidence frequency $\omega$ and vary $\omega_{p}$. In Figure~\ref{fig6}a,b,c we illustrate the reflection amplitudes of each harmonic with index $n$ when $\omega \sqrt{h/g}=0.1$ for increasing values of $\omega_{p}$. It is evident that the closer we move to the adiabatic limit where $\omega_{p}\ll \omega$ the better the agreement between the QS approximation and Floquet theory as a whole, while in all cases the harmonics at $n=-1,0,1$ are very well approximated by the QS approximation. A quantitative representation of the relative difference between the two methods, defined as $\delta|r_{n}|=||r_{n}|-|\tilde{r}_{n}||/|\tilde{r}_{n}|$, is depicted in Figure~\ref{fig6}d for $n=-2,-1,0,1,2$, spanning $\omega_{p}/\omega$ from 0 to 1. For $\omega_{p}\rightarrow 0$, we see that $\delta|r_{0}|\sim 10^{-7}$, $\delta|r_{\pm 1}|\sim 10^{-4}$ and $\delta|r_{\pm 2}|\sim 10^{-3}$, whereas for the case of panel (c) where the mismatch is more visible, $\delta|r_{0}|\sim 5 \cdot 10^{-4}$ and increases to $\delta|r_{\pm 2}|\sim 0.2$.

\begin{figure}[h]
\centering
\includegraphics[width=0.99\columnwidth]{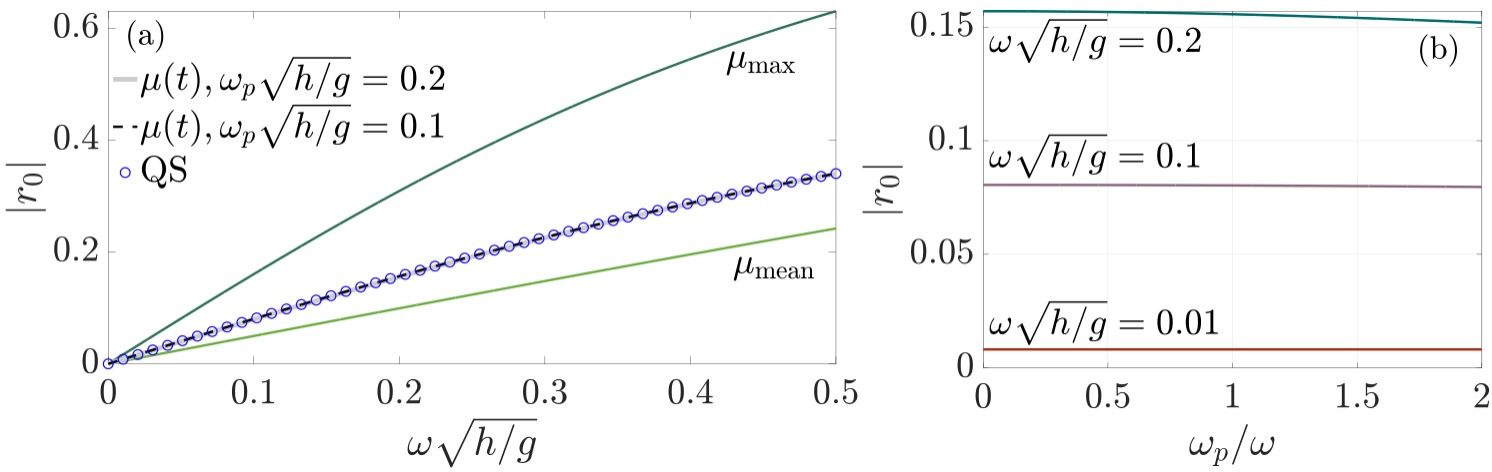}
\caption{(\textbf{a}) Comparison of the coefficient $|r_{0}|$ for a static plate of heights $\mu_{max}=0.95, \mu_{mean}=0.698$ with the oscillating plate of Figure~\ref{fig4}a where $\omega_{p}\sqrt{h/g}=0.2, 0.1$, along with the quasistatic result (QS) given by Equation \eqref{r_QS_explicit}. (\textbf{b}) Closer view of the change in $|r_{0}|$ with the normalized frequency of oscillation $\omega_{p}/\omega$ for $\omega\sqrt{h/g}=0.01, 0.1, 0.2$.\label{fig7}}
\end{figure}

Focusing on the fundamental frequency $n=0$, we wish now to inspect the effect of the plate when it is static versus when it is moving. Taking two limit values of $\mu$, specifically the maximum height $\mu_{max}=0.95$ and the mean value of $\mu(t)$, $\mu_{mean}=0.698$, we detect that $|R_{sw,\mu_{mean}}|<|r_{0}|<|R_{sw,\mu_{max}}|$ for all values of $\omega$ (see Figure~\ref{fig7}a). Interestingly, the coefficient $|r_{0}|$ is also minimally affected with changes of $\omega_p$, as already discussed in the Subsection~\ref{section_harmonic_generation}, and is almost perfectly captured by the result given from the quasistatic approximation. Arguably, a slight deviation from this quasistatic result can be achieved for higher values of $\omega$ and working towards fastest plate oscillations, i.e by challenging the limits of the shallow water approximation. This remark can be made by viewing Figure~\ref{fig7}b, where we find a relative variation of just $10^{-8}$ for $\omega \sqrt{h/g}=0.01$, of $10^{-2}$ for $\omega \sqrt{h/g}=0.1$ and of $3\cdot 10^{-2}$ for $\omega \sqrt{h/g}=0.2$ in an interval of $\omega_{p}/\omega \in [0,2]$. 
\section{Conclusion}
In this paper we proposed a Floquet theory approach for the reflection and transmission of a plane wave from a vertically oscillating plate in the low frequency approximation. With this shallow water approximation this oscillation of the plate is reduced to a 1D model with the simple wave equation and a time-varying point discontinuity condition at the plate location. We show that even a slowly oscillating plate can have an important impact on the generation of reflected harmonics, notably the first sidebands ($n=\pm1$) whose amplitude reaches close to $50\%$ the one of the fundamental one ($n=0$). When considering the quasistatic adiabatic limit $\omega_{p}\ll \omega$, explicit relations can be derived for the amplitudes of all harmonics which strongly agree with the full Floquet theory up to $n=\pm2$ for $\omega_{p}\rightarrow 0$. In addition, the reflection coefficient of the fundamental is nicely matched with the quasistatic result for all incident frequencies, and a slight variation when modifying $\omega_{p}$ can be only achieved by moving to higher frequencies and imposing even faster oscillations. This quasistatic value lies between the maximum and the mean value of the height oscillation. Overall, we conclude that the quasistatic approximation is surprisingly robust and can efficiently predict the behavior of the system with respect to the reflection coefficient of the fundamental harmonic regardless of the characteristic frequency of oscillation as long as the shallow water approximation is valid.

\end{document}